\documentclass{emulateapj}

\newcommand{\Msun}{{\rm M_{\odot}}}
\newcommand{\kpc}{\, {\rm kpc}}
\newcommand{\pc}{\, {\rm pc}}
\newcommand{\kmps}{\, {\rm km \, s^{-1}}}
\newcommand{\Tmb}{T_{\rm mb}}
\newcommand{\K}{\,{\rm K}}
\newcommand{\yr}{\,{\rm yr}}

\slugcomment{Manuscript in Preparation}
\shorttitle{Elongation and Supersonic Motions of Molecular Clouds}
\shortauthors{Koda et al}

\begin{document}

\title{The Elongations and Supersonic Motions of Molecular Clouds}


\author{Jin Koda\altaffilmark{1,2,3}, Tsuyoshi Sawada\altaffilmark{4}, Tetsuo Hasegawa\altaffilmark{3}, Nick Z. Scoville\altaffilmark{2}}


\altaffiltext{1}{JSPS Research Fellow: koda@astro.caltech.edu}
\altaffiltext{2}{California Institute of Technology, MS105-24, Pasadena, CA 91125}
\altaffiltext{3}{National Astronomical Observatory, Mitaka, Tokyo, 181-8588, Japan}
\altaffiltext{4}{Nobeyama Radio Observatory, Minamisaku, Nagano, 384-1305, Japan}


\begin{abstract}
New $^{13}$CO data from the BU-FCRAO Milky Way Galactic Ring Survey (GRS)
are analyzed to understand the shape and internal motions of molecular
clouds. For a sample of more than five hundred molecular clouds, we find
that they are preferentially elongated along the Galactic plane.
On the other hand, their spin axes are randomly oriented.
We therefore conclude that the elongation is not supported by internal
spin but by internal velocity anisotropy. It has been known that some
driving mechanisms are necessary to sustain
the supersonic velocity dispersion within molecular clouds.
The mechanism for generating the velocity dispersion must
also account for the preferred elongation. This excludes some driving
mechanisms, such as stellar winds and supernovae, because they do not
produce the systemic elongation along the Galactic plane.
Driving energy is more likely to come from large scale motions,
such as the Galactic rotation.
\end{abstract}


\section{Introduction}

The nature, origin, and maintenance of the supersonic motions or turbulence
in molecular clouds remain unresolved.
These motions must be 'continuously' driven, since otherwise they decay
on a cloud crossing timescale $\sim 3\times 10^6\yr$.
In fact, the free-fall collapse timescale of dense molecular clouds and
presumably the timescale for the decay of
supersonic motions are substantially shorter than the likely cloud lifetimes.
Such driving mechanisms must exert influence on the shape of molecular clouds.

There is an argument that molecular clouds are only converging regions
in interstellar turbulence, and thus, transient structures \citep{lar81,fle81}.
Even in this case, the shape of structures on scales similar to the cloud
sizes will reflect the mechanisms inputing energy at the top of the turbulent
cascade.
Using new data from the BU-FCRAO survey, we statistically investigate
the shape of molecular clouds in the inner Galaxy.

Possible energy sources for the interstellar turbulence have recently been
discussed on a theoretical basis \cite[see ][]{elm04, mac04}. The sources
may include stellar feedback (i.e. local events)
and galactic rotation (i.e. large-scale motions).
The physical mechanism converting these energies into turbulence has not
been well understood. We expect that it affects the morphology
of the gas structure as well as the turbulent motions.
In particular, the shape of molecular clouds might indicate the possible
energy source and physical mechanism. We will discuss the origin
of the velocity dispersions as well as the shape of molecular clouds.

\section{$^{13}$CO Molecular Clouds}

We use $^{13}$CO ($J=1-0$) line emission data from the BU-FCRAO Milky Way
Galactic Ring Survey \citep[GRS; see ][]{sim01}.
The data were downloaded from the GRS webpage in April 2004.
They cover almost the entire region in the Galactic longitude
$l \sim 16\arcdeg$ to $51\arcdeg$, latitude $b\sim -1.0\arcdeg$ to
$+0.5\arcdeg$, and velocity $-10$ to $130\kmps$.
The three-dimensional ($l$, $b$, $v$) data cube has a fully sampled
$22\arcsec$ grid in space and $0.22\kmps$ grid in velocity;
the resolutions are $46\arcsec$ (FWHM beam size) and $0.25\kmps$.


Molecular clouds are defined as topologically closed surfaces
with the peak main-beam brightness temperature $\Tmb$
above $4\K$ and the boundary temperature $2\K$ in the $^{13}$CO
data cube.
The $1\sigma$-noise level is $\Delta \Tmb = 0.48\K$ \citep{sim01},
and thus, emission features above $2\K$ ($4\sigma$) are all
significant.
The peak and boundary temperatures may be higher than those adopted
in past CO studies \citep[see ][]{sco87,sol87}.

The kinematic distances are used to estimate cloud sizes and hence
construct samples of small and large clouds.
Distance ambiguities are solved by utilizing the virial- and
LTE-mass relation as a distance indicator; we choose the distances
so that the virial- and LTE-masses become the most similar.
The relation is confirmed
for the clouds at terminal velocities, which do not suffer from
the distance ambiguity. The existence of the relation at
terminal velocities justifies the assumption that virtually all
clouds are gravitationally bound.

Each cloud is inspected by eye in $l-b$, $l-v$, and $b-v$ maps;
possibly-blended clouds (about 10\%) were removed from the sample
(note that this operation does not change the following results; and
the blending affected very few clouds as discussed in \S \ref{sec:spin}).
We also remove small clouds with diameter $D<5\pc$ in order to
select molecular clouds rather than small clumps and to
have clouds which were well-resolved. Past CO surveys had similar
cut-off for cloud identification because of their survey spacing
\citep[$3\arcmin$; ][]{sco87,sol87}.
The diameter is calculated as $D= (\Delta l  \Delta b)^{1/2}$ by using
the maximum linear extents in the $l$ and $b$ directions \citep{sco87}.
Incompletely sampled clouds at the edge of the cube are also removed.
The final catalog contains 552 clouds, most of which occupy more than
300 pixels each in the original data cube.

The data covers Galactic longitude $l=16-51\arcdeg$.
Its enclosed area is about $6.0\times 10^7\pc^2$, assuming $8.5\kpc$
for the Galactocentric radius to the Sun. 
The average surface number density of our $^{13}$CO clouds
is therefore $N_{\rm MC}\sim 10\kpc^{-2}$ in the inner Galaxy
($R_{\rm GC}\sim 4-8\kpc$). The number density is
$n_{\rm MC}\sim 100\kpc^{-3}$ assuming $100\pc$ for the thickness of
the Galactic molecular disk \citep{san84}.
These are slightly smaller than the values of CO molecular clouds
\citep[$15 \kpc^{-2}$, i.e. 1427 clouds within $9\times10^7\pc^2$; ][]{sco87}.

\begin{figure}
\epsscale{1.0}
\plotone{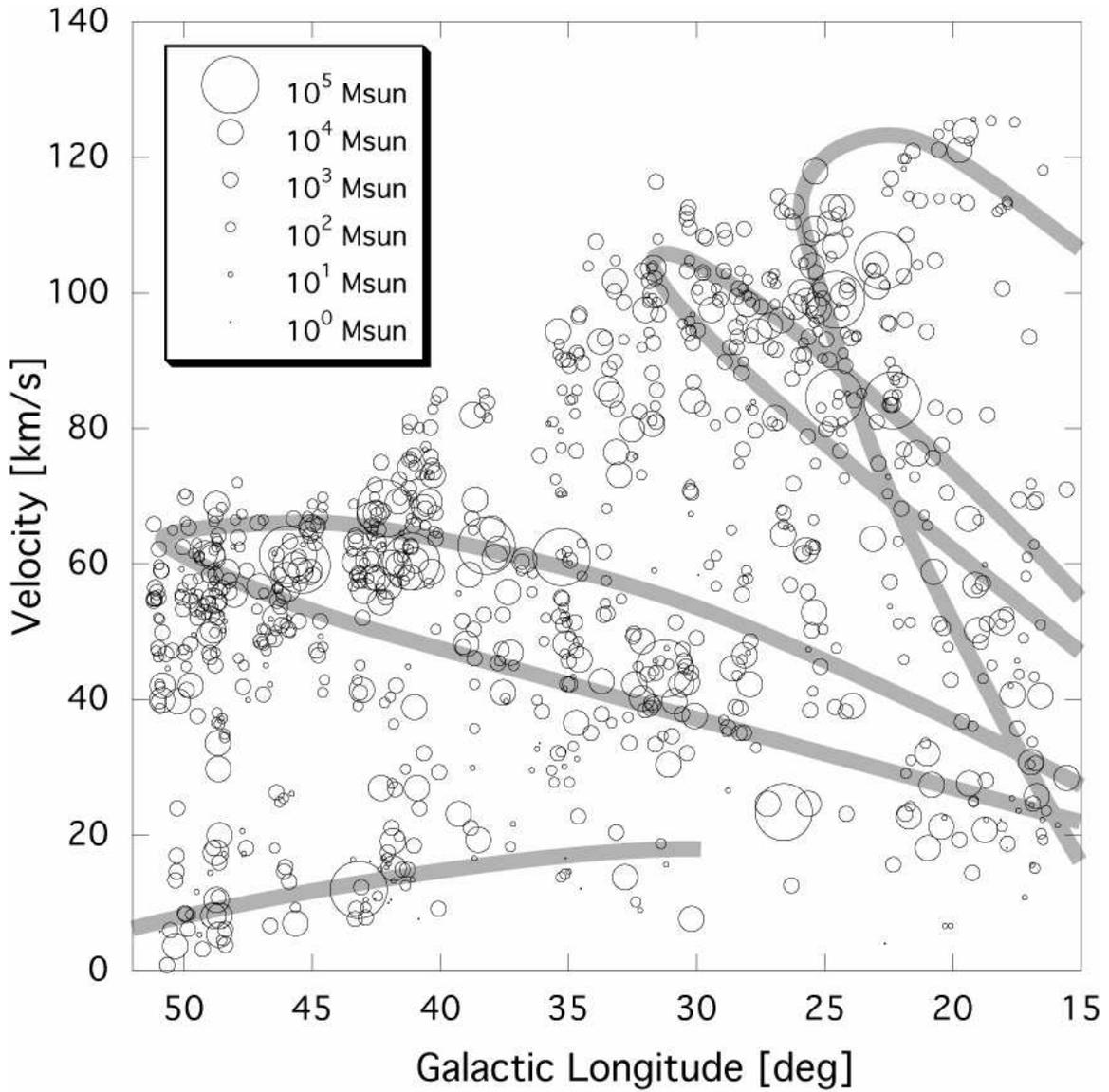}
\caption{$v-l$ diagram of the distribution of all identified
molecular clouds.
Thick grey lines indicate four major spiral arms, i.e. Perseus arm,
Sagittarius arm, Scutum arms, and $4\kpc$ arm (from left to right).
These spiral arms are identified entirely from radio recombination
observations, and not from CO observations (Downes et al. 1980;
see also Sanders et al. 1985).
\label{fig:lv}}
\end{figure}

Figure \ref{fig:lv} shows the distribution of molecular clouds in
$v-l$ diagram. Four thick lines indicate the Perseus arm,
Sagittarius arm, Scutum arm, and $4\kpc$ arm \citep[see ][]{san85}.
The other areas correspond to interarm regions. These spiral arms
are identified entirely from the distribution of HII regions observed
in radio recombination lines, and not from the molecular gas
distribution. {\it Most of massive $^{13}$CO
molecular clouds ($>10^5\Msun$) are associated with spiral arms,
but many less-massive clouds exist in interarm regions.}
This indicates either that the most massive molecular clouds are broken
up after passing spiral arms or that they cool down quickly so that
$^{13}$CO emission becomes undetectable. On the contrary, the
less-massive clouds survive across the spiral arms.

\section{Comparison with $^{12}$CO molecular clouds}

Table \ref{tab:mc} summarizes the average properties of $^{13}$CO
molecular clouds. We adopted two definitions of characteristic size
and velocity. The $D$ and $\Delta V$ are the diameter and full
velocity width of the clouds \citep{ss87}. The $S$ and $\sigma_{\rm V}$
are the intensity-weighted dispersions of size and velocity \citep{sol87}.
The virial mass was calculated with the equation in \citet{sol87}.

The diameter $D$ is consistent with the CO values both on the average
\citep[$18.1\pc$ in][]{sco87} and on the mass-weighted average
\citep[$40\pc$ in ][]{ss87}. Therefore, the extensions of $^{13}$CO
molecular clouds are almost the same as those of CO clouds.
The velocity dispersion $\sigma_{\rm V}$, however, is 1.4 times
smaller than the one in past CO studies \citep[$1.4\kmps$ in][]{sco87}.
The $\sigma_{\rm v}$-$D$ relation is $\sigma_{\rm v}=0.23 D^{0.54}$,
whose coefficient is $\sim1.4$ times smaller than the one in the
CO studies \citep[$\sigma_{\rm v} =0.31 D^{0.55}$ in][]{sco87}.
A natural interpretation may be that the central dense regions of
molecular clouds were saturated in the past CO studies. The $S$,
i.e. dispersion, is also smaller by a factor of several than the one
in CO studies \citep[][; see their Figure 1]{sol87}.
Further discussion of the size and mass distributions will be presented
in Sawada et al. (in preparation).

The averaged masses are different by an order of magnitude between
the mass-weighted average and normal average (Table \ref{tab:mc}).
This indicates that most molecular gas is in giant molecular clouds
($>10^5\Msun$), however, the majority of molecular clouds has only
$\sim 10^4\Msun$. These averaged masses are consistent with the ones
derived in the past CO studies \citep{sco87,ss87}, if the 1.4 times
difference of $\sigma_{\rm v}$ is taken into account.
The sound speed corresponding to the molecular gas temperature of $10\K$ is
$\sim 0.2\kmps$. The much larger velocity widths ($\Delta V$ and
$\sigma_{\rm V}$) indicate that the internal gas motions are supersonic.

\section{Elongation}

The position angles and axis ratios of molecular clouds are calculated
by diagonalizing the moment of inertia matrix as,
\begin{displaymath}
R_{-\theta}
\left(
\begin{array}{cc}
      \Sigma \, T_{ij} \alpha_{ij}^2     & - \Sigma \, T_{ij} \alpha_{ij} \beta_{ij}\\
    - \Sigma \, T_{ij} \alpha_{ij} \beta_{jj} & \Sigma \, T_{ij} \beta_{ij}^2
\end{array}
\right)
R_{\theta} =
\left(
\begin{array}{cc}
      I_{xx} & 0 \\
      0 & I_{yy}
\end{array}
\right),
\end{displaymath}
where $T_{ij}$ is the brightness temperature at a pixel ($i$,$j$),
and $\alpha_{ij}$ and $\beta_{ij}$ are the distances from the emission
centroid to the pixel in the $l$- and $b$-directions.
$R_{\theta}$ is a rotation matrix with the rotation angle $\theta$.
The axis ratio and position angle of clouds correspond to
$(I_{xx}/I_{yy})^{1/2}$ and $\theta$, respectively. The position
angle is defined from the positive $b$-direction going
counterclockwise to the positive $l$.
The direction along the Galactic plane is $\theta=90\arcdeg$.

\begin{figure}
\epsscale{1.0}
\plotone{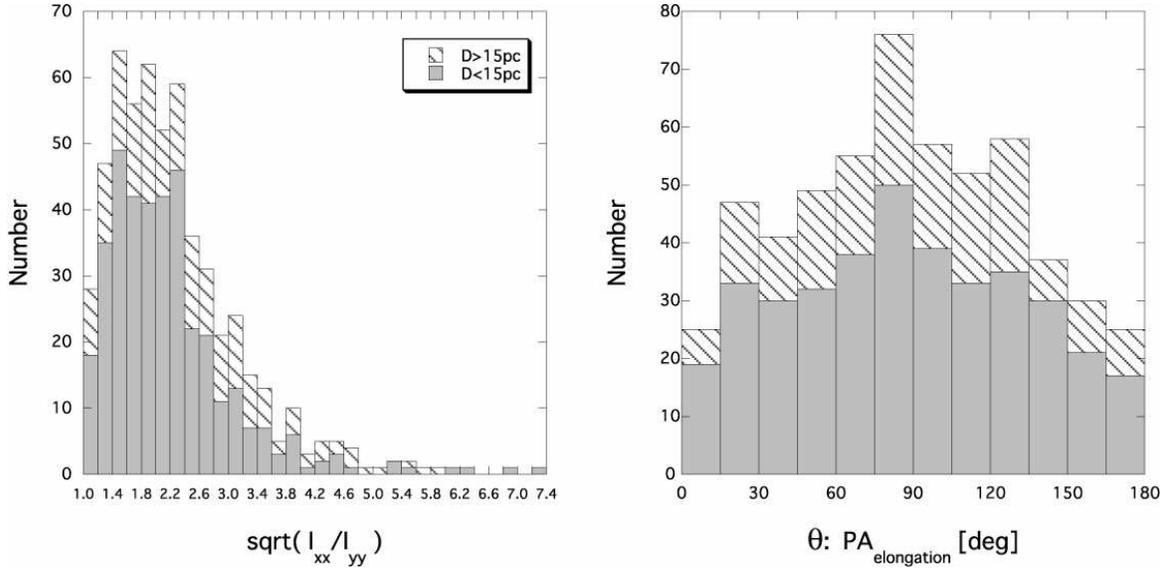}
\caption{Distribution of axis ratios ({\it left}) and position angles
({\it right}) of molecular clouds. The values are calculated using
the moments of inertia. Position angle $\theta$ is defined from the $+b$
direction going to positive $l$. Hence $\theta=90\arcdeg$ is the
direction along the Galactic plane. Large ($D>15\pc$) and small
($D<15\pc$) clouds are indicated with different patterns and
illustrated accumulatively.
\label{fig:elon}}
\end{figure}

Figure \ref{fig:elon} shows the histograms of axis ratio and
position angle. The axis ratio has the peak at about 1.8,
indicating that the molecular clouds are significantly elongated.
The population of round clouds, having the axis ratio of 1, is
small. The position angle peaks at $\theta \sim 90\arcdeg$ (i.e.
along the Galactic plane). Clouds with $\theta \sim 90\arcdeg$
is about two times more populated than those with $\theta
\sim 0\arcdeg$ or $180\arcdeg$.
{\it Therefore, molecular clouds are elongated
predominantly in the direction of the Galactic plane.}
We separate small clouds ($D<15\pc$) from large clouds ($D>15\pc$)
in Figure \ref{fig:elon}. As well as the large clouds,
the small clouds are predominantly elongated along the Galactic
plane. These clouds are much smaller than the width of the 
gas disk (about 100 pc) and stellar disk (about 600 pc).

\section{Spin}\label{sec:spin}
\begin{figure}
\epsscale{1.0}
\plotone{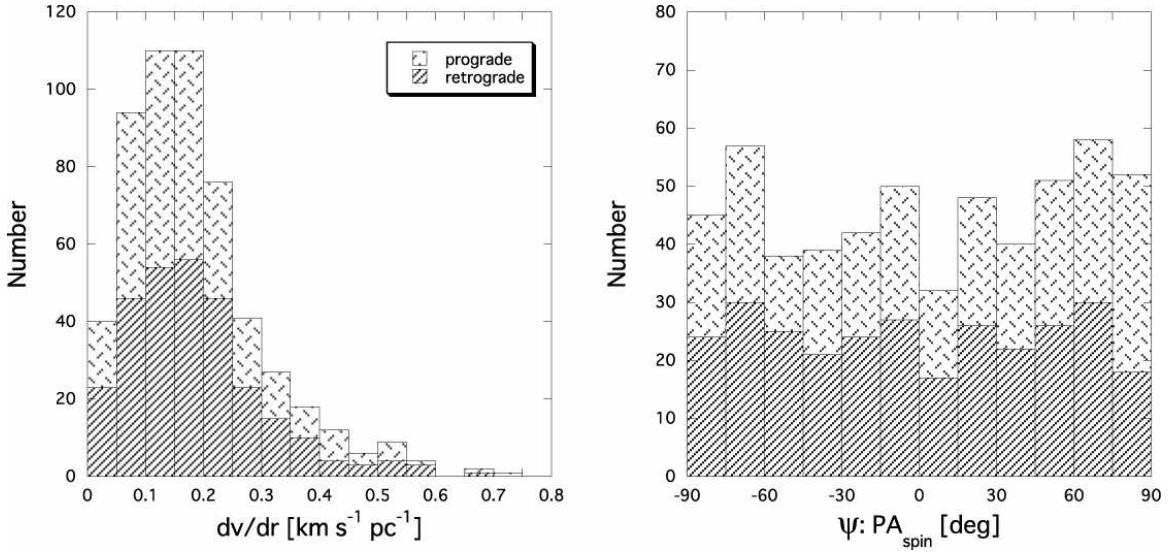}
\caption{
Distribution of velocity gradients ({\it left}) and position angles of
spin ({\it right}) of molecular clouds. Position angle $\psi$ is
defined as in Figure \ref{fig:elon} ({\it right}), but ranges between
$-90\arcdeg$ and $90\arcdeg$.  Hence, $\psi=0\arcdeg$ is the direction
perpendicular to the Galactic plane. The right figure can be compared
directly with Figure \ref{fig:elon} ({\it right}).
Prograde and retrograde spins are indicated with different patterns.
\label{fig:spin}}
\end{figure}

It is possible that the elongation is supported by the spin of
molecular clouds. In order to evaluate this, we measured
the velocity gradient in molecular clouds (projection on the sky)
and the position angle
perpendicular to the maximum velocity gradient, i. e. parallel to
the axis of possible spin (Figure \ref{fig:spin}).
The velocity gradient is estimated from the observed velocity field
by fitting a plane.
The position angle $\psi$ is oriented with $0\arcdeg$ at $+b$,
increasing toward $+l$.
Typical velocity gradients $dv/dr$ projected on the sky are
about 0.15 $\kmps\,\pc^{-1}$. We did not correct for the apparent
velocity gradient due to the projected LSR velocity over the
diameter of the clouds, since this gradient is at most $\sim
0.04\kmps\,\pc^{-1}$, assuming a rotation velocity of $220\kmps$.
Prograde and retrograde spins with respect to the Galactic rotation
are indicated with different patterns, and are equally populated.

\begin{figure}
\epsscale{0.5}
\plotone{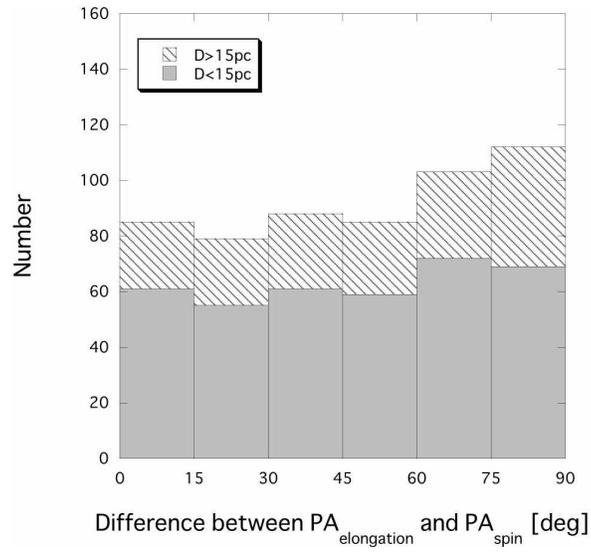}
\caption{
Difference between the position angles of elongation and spin.
\label{fig:diff}}
\end{figure}

Figure \ref{fig:spin} {\it right} can be directly compared to
Figure \ref{fig:elon} {\it right}.
If spin supports the elongation, the $\psi$ should be distributed
similarly to the $\theta$; their peaks should be related as
$\psi_{\rm peak} \sim \theta_{\rm peak}-90\arcdeg$. The figures, however,
show that the spin is randomly oriented and has no dominant peak.
Figure \ref{fig:diff} shows the difference of the position angles of
the elongation axis and spin axis. This figure shows no significant
peak at $90\arcdeg$, compared with Figure \ref{fig:elon} {\it right}
and considering the statistical error $\sim N^{1/2}$.
{\it Therefore, the cloud elongations are not
supported by internal spins, but should be related to the internal
velocity dispersions (supersonic motions).}


It is noteworthy that confusion (blending) in cloud identification
is likely to align the apparent spin axis perpendicular to an elongation
-- it is most likely that two clouds are slightly offset in space and
velocity. The opposing result strongly indicates that
the blending is not important in our sample.

\section{Discussion}

The supersonic motions within molecular clouds must be regenerated
continuously; otherwise they decay rapidly within a cloud crossing
time. {\it Mechanisms for driving the supersonic motion
must also account for the cloud elongation along the Galactic
plane.} 
In light of the new results, we discuss a possible energy source and
mechanism for driving supersonic motions and cloud elongations.

\subsection{Energy Source}

Among several suggested energy sources \citep[see ][]{elm04},
two sources, i.e. stellar feedback and galactic rotation,
are most often considered.
Most other sources extract their intrinsic energies from
these two.

Stellar feedback includes protostellar outflows, stellar
winds, ionizing radiation, and supernova explosions.
These feedback mechanisms, however, release energy isotropically
or in random directions. Such processes are unlikely to cause
the preferential orientation of molecular clouds.
They would cause a random orientation.

Supernovae are the most energetic feedback mechanism
\citep{nor96,mac04}. There are
some arguments for supernovae-induced dense gas formation and
supersonic turbulence. If supernova blastwaves, however, compress
surrounding material and induce molecular cloud formation,
their round expansion will produce {\it no} preferential orientation.
The blastwaves would escape from the Galactic disk predominantly
in the direction normal to the Galactic plane. Then, most molecular
clouds would be elongated perpendicular to the plane. The stellar
feedback is unlikely to be the dominant cause of the elongation
of molecular clouds.

Evidence has been accumulated for the presence of molecular clouds
without star formation activity \citep{moo88,sco89,wil98}.
Stellar feedback cannot supply energy to such clouds. High mass
star-forming regions are localized around spiral arms in the Galaxy
\citep{dow80, san85} and in external galaxies \citep{sco01}.
It is also unlikely that the energy supply from stellar feedback
suffices to maintain supersonic motions in interarm molecular clouds.

Since the cloud elongation has the preferential direction,
the energy source is likely to have a similar preferred direction.
Galactic rotation has an rotation axis, and is a likely energy
source. Galactic rotation provides an enormous reservoir of
energy; it can maintain the elongations and supersonic motions
over the Galactic age.

Even if molecular clouds are only transient structures
in large-scale interstellar turbulence, the anisotropic shapes should
indicate the origin of the turbulence. It is still likely that the
Galactic rotation is the main source of the turbulant energy.

\subsection{Driving Mechanism}\label{sec:mec}
How might the rotation energy be transfered to random motion in
molecular clouds? Energy transfer from rotation to random motions
generally occurs in a differentially rotating disk, if mass elements
orbiting at slightly different radii pull on each other \citep{sel99}.
\citet{sel99} pointed out that many types of viscous stress, including
Reynolds, Maxwell, and gravitational stresses, can cause the energy
transfer from rotation to random motion.

For example, \citet{jog88} found with an analytic calculation that
cloud-cloud gravitational interactions (gravitational stress) can
transfer the orbital energy of clouds into random motions
{\it between} clouds [but they did not discuss internal motions].
\citet{wad02} used hydrodynamical simulations and
showed that the rotational energy could cascade into smaller scales
(down to their resolution limit, i.e. the size of giant molecular
clouds), owing to local
tidal force among clouds, filaments, and voids.
Perhaps, this cascade proceeds down to the internal motions of
the clouds.

If tidal gravitational torques exerted by passing clouds is
the cause of the velocity dispersions, the preferred
elongation of molecular clouds would be produced naturally.
The mean separation of molecular clouds is about
$1/n_{\rm MC}^{1/3} \sim 200\pc$. This is larger than
the thickness of the Galactic molecular disk $100\pc$.
Tidal radius of molecular clouds should be at least a few
times larger than their diameter,
and thus is not negligible compared with the disk thickness.
Tidal interaction should occur
predominantly in the Galactic plane; it may generate asymmetric
velocity dispersions which cause the elongation along the Galactic
disk.

A slight exchange of angular momentum may occur during tidal
interaction and spin up clouds. The spin axis might become
perpendicular to the elongation, however, this effect should be
little. In case that two clouds pass by at the distance $d=100\pc$
and velocity $v=6\kmps$ (typical cloud-cloud velocity dispersion),
the tidal interaction would increase the velocity gradient within a
cloud by at most $v/2\pi d \sim 0.01\kmps\,\pc^{-1}$ if the two
clouds are strongly coupled. This is negligible compared
with the observed velocity gradient.
In addition, random encounters would suppress the increase of
angular momentum. The spin-up is not significant even if the tidal
interaction is the cause of supersonic motions.

We discussed only gravitational interactions between molecular clouds.
Some other density fluctuations, such as filamentary structure in gas,
would also strengthen the tidal field around clouds.
There are, of course, some other possible mechanisms introduced by
magnetic field, surrounding diffuse gas (HI and $\rm H_2$),
and gradients in stellar potential. Further discussions would require
theoretical researches.

\section{Summary}

We showed that $^{13}$CO molecular clouds are elongated predominantly
along the Galactic plane. The elongations are not supported by the
spins of molecular clouds, but should be supported by supersonic
velocity dispersions in molecular clouds. In order to avoid a rapid
decay of the supersonic motions, there must be some mechanisms to
drive them continuously. The driving mechanisms must
also account for the elongations of molecular clouds. The driving
energy is likely to come from large-scale motions, such as the
Galactic rotation, and not from stellar feedback.

\acknowledgments
The Boston University-FCRAO Milky Way Galactic Ring Survey (GRS) is
a joint project of Boston University and Five College Radio Astronomy
Observatory, funded by the National Science Foundation under grants
AST-9800334, AST-0098562, \& AST-0100793.
We thank Keiichi Wada, Masahiro Sugimoto, and Toshihiro Handa for
useful discussions. JK was financially supported by the Japan Society
for the Promotion of Science (JSPS) for Young Scientists. This work
has been partially supported by National Science Foundation under grant
9981546.


\clearpage

\begin{table}
\begin{center}
\caption{Properties of $^{13}$CO Molecular Cloud\label{tab:mc}}
\begin{tabular}{lcc}
\tableline \tableline
Parameter & mass-weighted mean & mean \\
\tableline
$D$ (pc)$^a$   & 38 & 15 \\
$S$ (pc)$^b$ & 8 & 3 \\
$\Delta V$ ($\kmps$)$^c$  & 11.5 & 5.1 \\
$\sigma_{\rm v}$ ($\kmps$)$^d$ & $2.2$ & $1.0$ \\
$\rm T_{peak}$ (K)$^e$     & $11$ & $8.9$ \\
$\rm M_{VT}$ ($\Msun$)$^f$  & $1.2\times10^5$ & $1.3\times10^4$ \\
\tableline
\end{tabular}
\tablecomments{Mean parameters are calculated among molecular clouds
with $D>5\pc$}
\tablenotetext{a}{Diameter of molecular cloud}
\tablenotetext{b}{Intensity-weighted size}
\tablenotetext{c}{Velocity width at the 2$\K$ level}
\tablenotetext{d}{Intensity-weighted velocity dispersion}
\tablenotetext{e}{Excitation temperature (calculated from main beam
temperature and cosmic microwave background temperature (2.7 K))}
\tablenotetext{f}{Virial mass}
\end{center}
\end{table}

\end{document}